# Optimum Design of a 4x4 Planar Butler Matrix Array for WLAN Application

Wriddhi Bhowmik, Shweta Srivastava

**Abstract**— In recent years, high-speed wireless communication is in vogue. In wireless communication systems, multipath fading, delay and interferenc occurres by reflection or diffraction. In a high-speed wireless communication, it becomes a necessary to separate desired signal from delay or interference signal. Thus to overcome these problems Smart antenna systems have been developed. Basically there are two types of smart antenna systems, one is Switched beam system and another Adaptive array system. This paper presents the optimum design of a 4x4 plannar Butler matrix array as a key component of a switched beam smart antenna system, operating at 5.2 GHz for WLAN with a dielectric substrate, FR4 of $\varepsilon_r$ =4.9 and h=1.6mm. Conception details, simulation results and measurements are also given for the components (microstrip antenna, hybrid couplers, cross-coupler, phase shifter) used to implement the matrix. In this dissertation, mathematical calculations for all the components using MATLAB is done and then every individual component is designed using the commercial software SONNET. Then these entire components have been combined on a single substrate and simulated using SONNET.

**Index Terms**—Switched beam system, 4x4 Butler matrix array, Microstrip.

———————————— ◆ ————————————

## 1 INTRODUCTION

Now Smart antenna [1] is one of the most promising technologies that will enable a higher capacity in wireless networks by effectively reducing multipath and co-channel interference. This is achieved by focusing the radiation only in the desired direction and adjusting itself to changing traffic conditions or signal environments. Smart antennas employ a set of radiating elements arranged in the form of an array.

Switched beam systems [2] are referred as antenna-array systems that form multiple-fixed beams [3] [4] with enhanced sensitivity in a specific area. This antenna system detects signal strength, selects one of the several predetermined fixed beams, and switches from one beam to another as the user moves.

This paper introduces the 4x4 planar Butler matrix array [2] [3] [4] as a key component of the switched beam system. Basically this Butler matrix array forms multiple fixed overlapping beams which will cover the designated angular area. It is a NxN passive feeding network with N radiating elements. The output ports of the butler matrix feed the antenna elements. It is easy to implement and requires few components to build compared to the other networks. The loss involved is very small, which comes from the insertion loss in hybrids, phase shifters and transmission lines. However in a butler matrix, beamwidth and bean angles tend to vary with frequency. Also it has a complex interconnection scheme for a large matrix.

Butler matrix has been implemented with various techniques such as waveguide, microstrip, multilayer microstrip, suspended stripline, CPW etc. Microstrip technique is widely

————————————————
- Wriddhi Bhowmik is with the Department of ECE, Birla Institute of Technology, Mesra, Ranchi, India.
- Dr. Shweta Srivastava is with the Department of ECE, Birla Institute of Technology, Mesra, Ranchi, India.

used in Butler matrix due to its numerous advantages such as low profile, easy fabrication and low cost.

In this paper we are using the microstrip technique [5] [6] to implement the Butler matrix array. The major components of the butler matrix such as hybrid couplers, crossovers and phase shifters are designed by microstrip line technique and the array is formed by four rectangular patch microstrip antennas with inset feeding. Antennas provide the directional pattern individually.

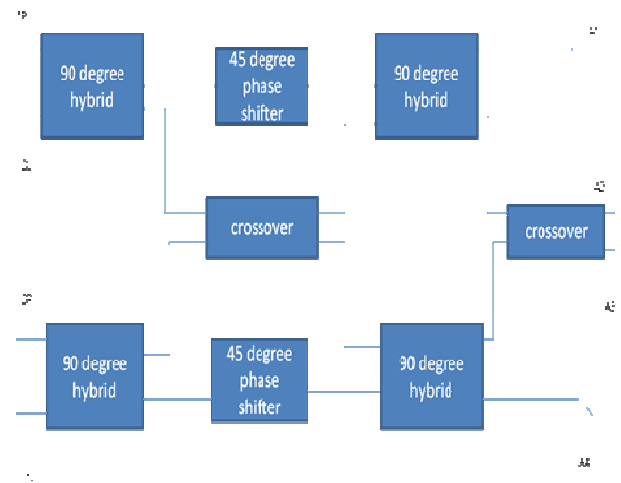

Fig 1: Block diagram of 4x4 Butler matrix array

As all the radiating elements are implemented together the system can provide narrow beams in different directions with higher gain. The microstrip technique is used to implement the matrix as well as the antenna array on the same substrate.

This paper presents the optimum design and the simulation results of 4x4 planar Butler matrix array and all its individual components.

## 2 ANALYSIS AND DESIGN

The butler matrix [2] [7] [9] is a NxN network consisting of N input ports and N output ports, $(N/2)\log_2(N)$ $90^0$ hybrids and $(N/2)(\log_2(N)-1)$ fixed phase shifters to form the beam pattern.

The NxN Butler matrix creates a set of n orthogonal beams in space by processing the signal from N antenna elements. These beams are pointing in the direction θ governed by the following equation:

$$\sin\theta_i = \pm(i\lambda/2Nd) \quad \ldots\ldots\ldots\ldots (1)$$

Where i = 1, 2, 3,…,(N-1). The corresponding inter element phase shift with spacing d=λ/2 is

$$\alpha_i = \beta d \sin\theta_i = i(\pi/N) \quad \ldots\ldots\ldots(2)$$

Where $\beta = 2\pi/\lambda$, is the wave number. This paper presents the optimum design of 4x4 planar Butler matrix array which consist of four $90^0$ hybrid, two crossovers and two $-45^0$ phase shifters. From fig 1 it is evident that the Butler matrix has four inputs 1R, 2L, 2R, 1L and four outputs A1, A2, A3, A4. These four outputs are used as inputs to antenna elements to produce four beams. The input ports of the Butler matrix are named according to their beam position.

### 2.1 $90^0$ Hybrid

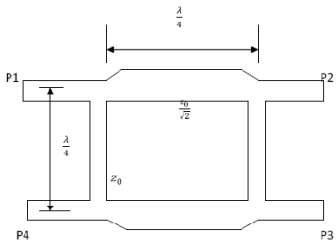

Fig 2: Geometry of $90^0$ hybrid

Quadrature hybrid [5] [6] or $90^0$ hybrid is a well known device used for its ability to generate signals 90 degree out of phase at its outputs. Through Even-Odd mode analysis it can be shown that the S matrix will have the following form,

$$S = -\frac{1}{\sqrt{2}}\begin{bmatrix} 0 & j & 1 & 0 \\ j & 0 & 0 & 1 \\ 1 & 0 & 0 & j \\ 0 & 1 & j & 0 \end{bmatrix}$$

From the above figure it is evident that the quadrature hybrid has four port, port 1 is the input port, port 2 is the output port, port 3 is the coupled port and port 4 is the isolated port. When power is applied to the port 1 it is equally distributed in port 2 and port 3 and port 4 is isolated since no power reaches it. There is a $90^0$ phase difference between port 2 and port 3.

### 2.2 CROSSOVER

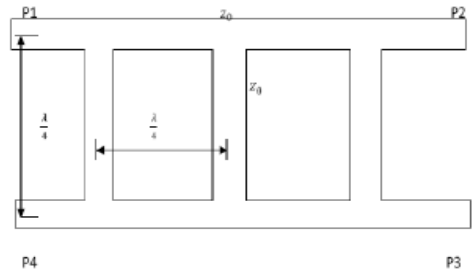

Fig 3: Geometry of the crossover

By cascading two $90^0$ hybrids we can implement a crossover but it does not give satisfactory performance so the geometry used to implement the crossover [10] is shown in fig3. The S matrix for the crossover can be obtained by extending the even odd mode analysis for the quadrature hybrid. The S matrix will have the following form:

$$S = \begin{bmatrix} 0 & 0 & j & 0 \\ 0 & 0 & 0 & j \\ j & 0 & 0 & 0 \\ 0 & j & 0 & 0 \end{bmatrix}$$

### 2.3 PHASESHIFTER

The phase shifter [5] [6] is implemented using microstrip transmission line. The length of the line corresponding to $45^0$ phase shift is given by the formula

$$\Phi = (2\pi/\lambda)L \quad \ldots\ldots\ldots\ldots(3)$$

Where L is in meters, φ is in radians. λ is the wavelength in the microstrip line. The wavelength in the microstrip transmission line is given by

$$\lambda = \lambda_0/(\epsilon_{reff})^{0.5} \ldots\ldots..\ldots(4)$$

Where $\lambda_0$ is the free space wavelength and $\epsilon_{reff}$ is the effective dielectric constant of the microstrip line. Since the phase shift is implemented using simple transmission line therefore it is linearly frequency dependent.

## 3 EXPERIMENTAL SETUP AND RESULTS

At first the important mathematical calculations for finding the dimensions of all the individual components is done using MATLAB and then the individual components are designed and simulated using SONNET. After getting good results all the individual components are combined on a single substrate to implement the Butler matrix and simulated using SONNET.

### 3.1 90⁰ Hybrid

90⁰ hybrid is made by two main transmission lines shunt connected by the two secondary branch lines. It has two 50Ω and two 35.4 Ω transmission lines with length λ/4. So the perimeter of the square is approximately equal to one wavelength. Some mathematical equations used to calculate the width and the length of the transmission lines are given below, for $Z_0 \leq 44 - 2\epsilon_r$

$$W/h = \frac{2}{\pi}\{B-1-\log(2B-1)+\frac{\varepsilon_r-1}{2\varepsilon_r}[\log(B-1)+0.39-\frac{0.61}{\varepsilon_r}]\} \quad \ldots\ldots\ldots(5)$$

otherwise

$$W/h = \frac{8e^A}{e^{2A}-2} \quad \ldots\ldots\ldots(6)$$

where

$$A = z_0(\varepsilon_r + 1/2)^{0.5} + \frac{\varepsilon_r-1}{\varepsilon_r+1}(0.23+\frac{0.11}{\varepsilon_r}) \quad \ldots\ldots\ldots(7)$$

and

$$B = \frac{60\pi^2}{z_0\sqrt{\varepsilon_r}} \quad \ldots\ldots\ldots(8)$$

W is the width of microstrip lines [5] [6] and h is the height of the substrate FR4 and the length of the microstrip line is calculted by the following formula

$$L = \frac{c}{4f\sqrt{\varepsilon_{reff}}} \quad \ldots\ldots\ldots(9)$$

where f is the operating frequency, c is speed of light in free space and $\varepsilon_r$ is the effective dielectric constant. Now the optimum design of 90⁰ hybrid obtained by the SONNET is given below

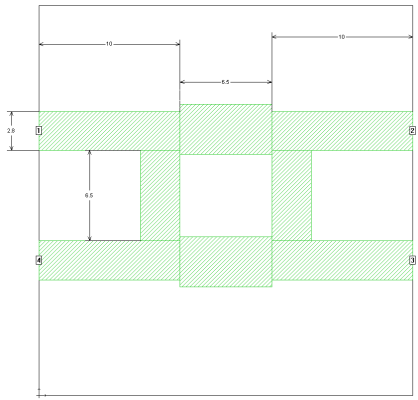

Fig 3: Optimum design of 90⁰ hybrid at 5.2GHz

| Parameters | Calculated values | Measured values |
|---|---|---|
| Width of 50 ohm line | 2.8 mm | 2.8 mm |
| Width of 35.4 ohm line | 4.8 mm | 3.8 mm |
| Length of the line | 7.5 mm | 6.5 mm |

Simulation results are given below

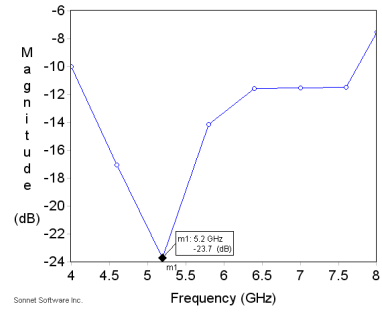

Fig 4: Returnloss vs frequency, $s_{11}$= -23.7dB

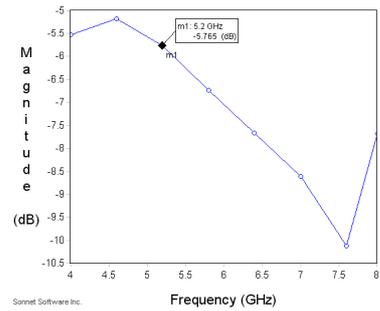

Fig 5: Coupling level vs frequency, $s_{13}$= -5.7dB

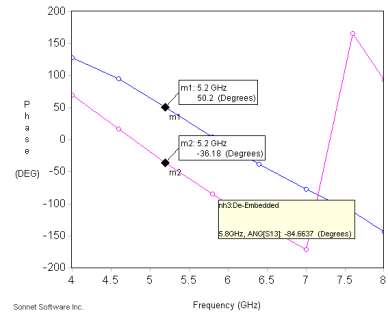

Fig 6: Phase difference = $s_{12} - s_{13}$ = 86.4⁰

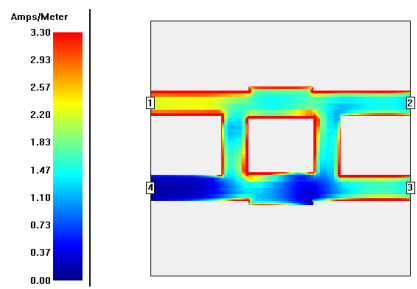

Fig 7: Current distribution at 5.2GHz

### 3.2 Crossover
The crossover is implemented by 50Ω microstrip line. Optimum design is given below

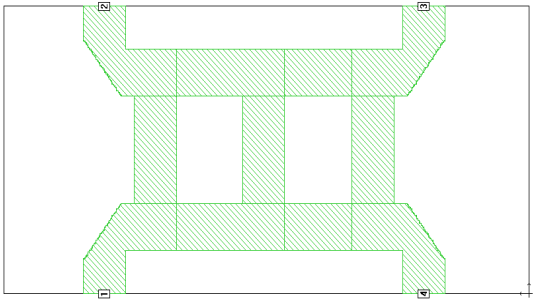

Fig 8: Optimum design of crossover at 5.2GHz.

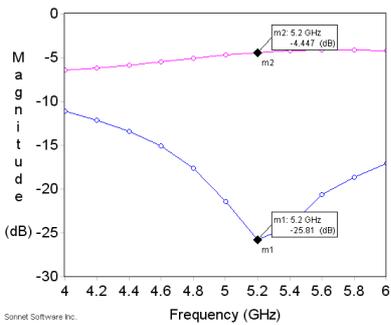

Fig 9: returnloss $S_{11}$= -25.81dB and coupling $S_{13}$= -4.45dB.

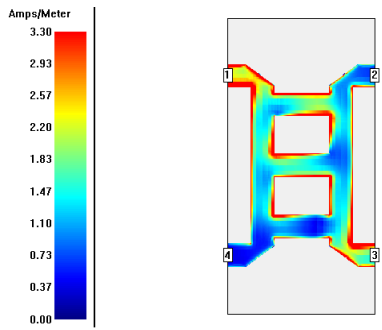

Fig 10: Current distribution at 5.2GHz.

### 3.3 Phaseshifter
In this paper $-45^0$ phaseshifter [5] is used. The phaseshifter is implemented using the microstrip lines and the length of the microstrip lines are calculated by the eq(3) and eq(4). Optimum design is given below

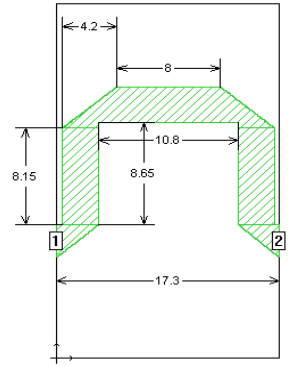

Fig 11: Optimum design of phaseshifter at 5.2GHz.

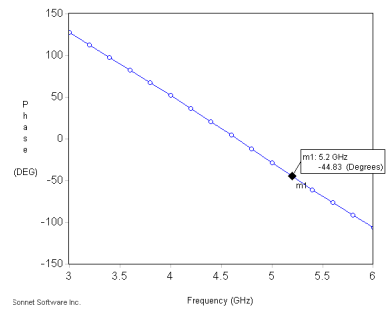

Fig 12: phaseshiftining = $-44.83^0$.

### 3.4 Rectangular patch microstrip antenna
Output ports of the butler matrix feed the antenna elements. The 4x4 butler matrix feed 4 radiating elements, so we can get the fixed overlapping beams directed in the different directions and cover the $120^0$ cellular area. Here rectangular patch microstrip antenna [8] is used as a radiating element. Width and length of the rectangular patch and the length of the inset feed is calculated by the following formulae

$$W = \frac{v_0}{2f_r}\sqrt{\frac{2}{\varepsilon_r+1}} \quad \ldots\ldots\ldots\ldots\ldots\ldots\ldots\ldots(10)$$

$$\varepsilon_{reff} = \frac{\varepsilon_r+1}{2} + \frac{\varepsilon_r-1}{2}[1 + 12\frac{h}{w}]^{(-0.5)} \quad \ldots\ldots\ldots\ldots(11)$$

So,

$$\frac{\Delta L}{h} = 0.412 \frac{(\varepsilon_{reff}+0.3)(\frac{w}{h}+0.264)}{(\varepsilon_{reff}-0.258)(\frac{w}{h}+0.8)} \quad \ldots\ldots\ldots\ldots(12)$$

$$L = \frac{v_0}{2f_r\sqrt{\varepsilon_{reff}}} - 2\Delta L \quad \ldots\ldots\ldots\ldots\ldots(13)$$

and the location of inset feeding is derived by the following formula

$$R_{in}(y=y_0) = R_{in}(y=0)\{\cos(\frac{\pi}{L}y_0)\}^2 \quad \ldots\ldots(14)$$

W is the width of the patch and L is the length, the length of the patch has beed extended along with its length on each side by a length ΔL due to the fringing effect and h is the height of the substrate. Optimum design is given below

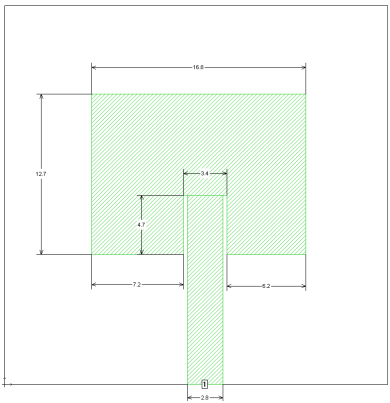

Fig 13: Optimum design of rectangular patch antenna at 5.2GHz.

| Width of antenna | 16.8 mm |
|---|---|
| Length of antenna | 12.7 mm |
| Inset feeding length | 4.7 mm |
| Spacing | 3.4 mm |
| Width of microstrip line | 2.8 mm |

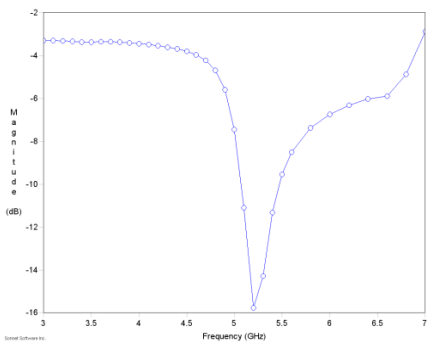

Fig 14: returnloss $S_{11}$ = -15.8dB at 5.2GHz.

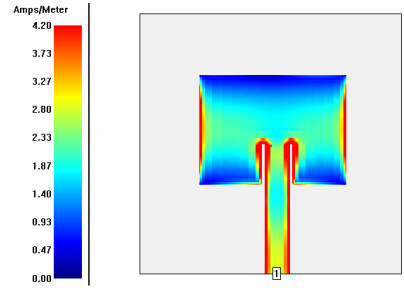

Fig 15: Current distribution at 5.2GHz.

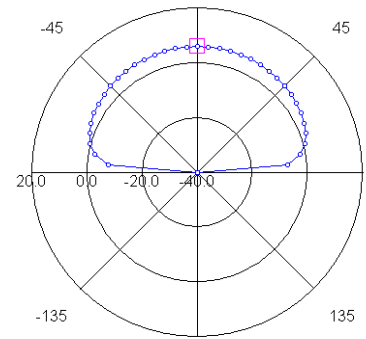

Fig 16: Radiation pattern at 5.2GHz.

## 3.5 Design of 4x4 Butler matrix array

Now all the individual components are combined on a single substrate FR4 to implement the butler matrix array [4] [9] [11]. Optimum design of butler matrix array is given below

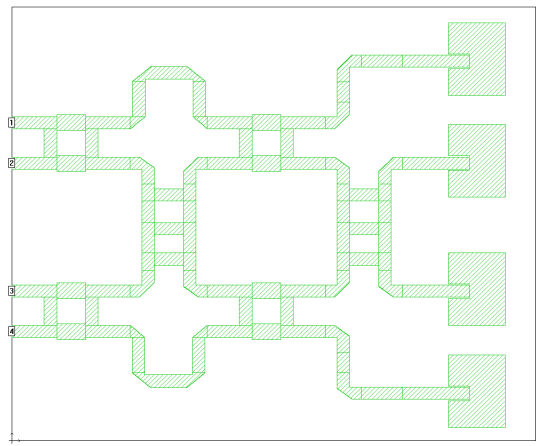

Fig 17: Optimum design of 4x4 butler matrix array at 5.2GHz.

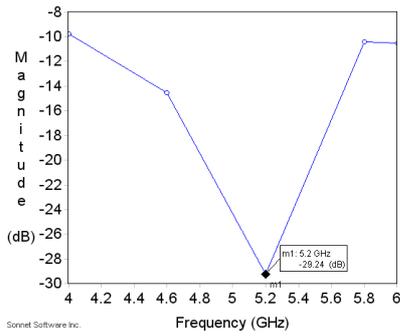

Fig 18: returnloss $s_{11}$= -29.24dB

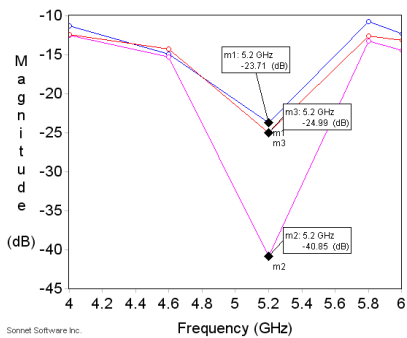

Fig 19: returnlosses for different ports, $s_{22}$= -23.71dB, $s_{33}$= -25dB, $s_{44}$= -40.85dB.

The coupling levels [2] of the butler matrix array is given below when the different ports are fed

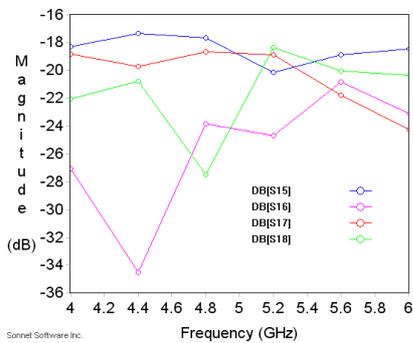

Fig 20: coupling levels $S_{15}$, $S_{16}$, $S_{17}$, $S_{18}$ at 5.2GHz.

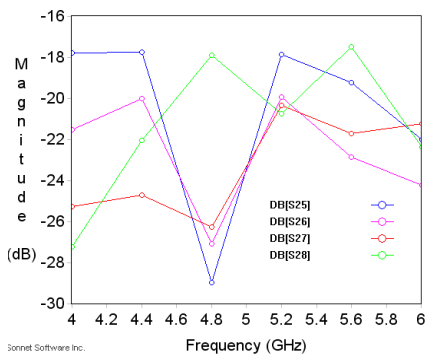

Fig 21: coupling levels $S_{25}$, $S_{26}$, $S_{27}$, $S_{28}$ at 5.2GHz.

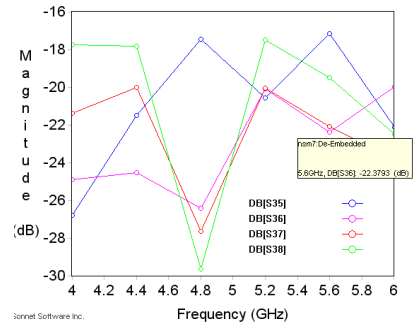

Fig 22: coupling levels $S_{35}$, $S_{36}$, $S_{37}$, $S_{38}$ at 5.2GHz.

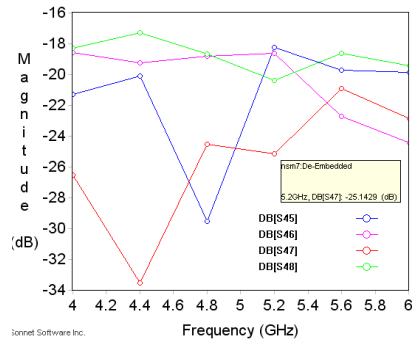

Fig 23: coupling levels $S_{45}$, $S_{46}$, $S_{47}$, $S_{48}$ at 5.2GHz.

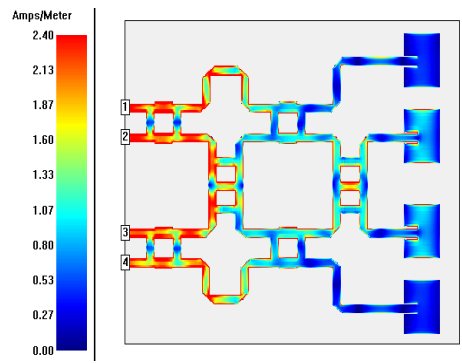

Fig 24: Current distribution when all the ports are fed at 5.2 GHz.

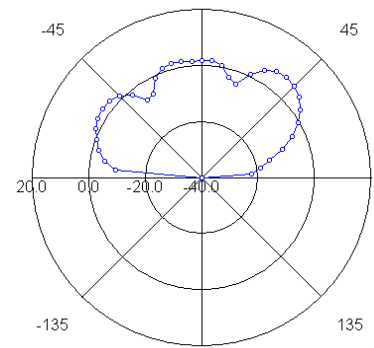

Fig 25: Radiation pattern of 4x4 butler matrix array when all ports are fed at 5.2GHz.

74

## 4. CONCLUSION

This paper presents the optimum design of a 4x4 planar Butler matrix array for WLAN application. Initially all the necessary formulae for dimensions of hybrid coupler, crossover, phase-shifter and radiating elements are evaluated using MATLAB. Then all the components are designed and simulated using commercial software SONNET. After achieving proper simulation results of all individual components, these components are then combined on a single substrate and simulated using SONNET. Here a high dielectric constant is considered for the design, hence after studying the simulation results of individual components and the whole structure it is evident that high loss has occurred. To reduce this loss a low dielectric constant can be used. This design will cover $120^0$ cellular area, for larger coverage the design should be extended to 8x8 matrix. Also to increase the gain, different kinds of radiating elements should be used. These designs need to be verified experimentally.

## REFERENCES

[1] Jack Winters, "Smart Antennas for Wireless Systems," *IEEE Personal Communications, pp. 23-27, Feb. 1998.*

[2] "Switched Beam Antenna using Omnidirectional Antenna Array" by Siti Zuraidah Ibrahim and Mohamad Kamal A.Rahim , Asia-Pacific Conference on Applied Electromagnetics Proceedings 2007, Malaysia.

[3] Siti Rohaini Ahmad and Fauziahanim Che Seman, "4-Port Butler Matrix for Switched Multibeam Antenna Array" *IEEE, Asia-Pacific Conference on Applied Electromagnetics Proceedings, 2005, Johor, Malaysia, December 20-21.*

[4] D. C. Chang, S. H. Jou. 'The study of Butler Matrix BFN for four beams antenna system', *IEEE Antenna and Propagation Society International Symposium, volume 4, pp. 176-179, (2003).*

[5] Pozar D.M: "Microwave Engineering", Third edition, Wiley, 2005.

[6] Fooks, E. H. Microwave engineering using microstrip circuits, Prentice Hall New York 1990.

[7] R. Comitangelo, D. Minervini, B. Piovano, "Beam Forming Networks of Optimum Size and Compactness for Multibeam Antennas at 900 MHz", *IEEE Antenna and Propagation International Symposium, Vol. 4, pp. 2127-2130, July, 1997.*

[8] Constantine A. Balanis, 'Antenna theory analysis and design', second edition, John willey and sons, Inc., 1997.

[9] Tayeb. A. Denidni and Taro Eric Libar, "Wide Band Four-Port Butler Matrix for Switched Multibeam Antenna Arrays," *The 14th IEEE 2003 International Symposium on Personal, Indoor and Mobile Radio Communication Proceedings, pp.2461-2464, 2003.*

[10] Wight, J S., chudobiak W. J., "The microstrip and stripline crossover structure" *IEEE trans. On microwave theory and techniques, May 1976, p-270.*

[11] Nhi T. Pham, Gye-An Lee and Franco De Flaviis, *Microstrip Antenna Array with Beamforming Network for WLAN Application* , Department of Electrical and Computer science, University of California,USA.
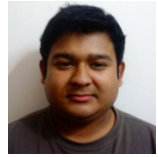

**Mr. Wriddhi Bhowmik** received his B.Tech in 2008 from West Bengal University of Technology, Kolkata, India. Presently he is pursuing his M.E from Birla Institute of Technology Mesra, Ranchi, India, in the field of Wireless Communication. His areas of interest are Wireless Communication and Antennas.

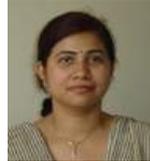

**Dr. Shweta Srivastava** received her Ph.D. from BHU, Varanasi (U.P.)-India. Presently she is working as senior lecturer in the Department Electronics & Communication Engg., B.I.T, Mesra, Ranchi, India. She is currently supervising a project named "Design of Active Microstrip Antenna for Wireless Communication" funded by DST, SERC Fast Track Scheme for Young Scientists. Her areas of interest are Microstrip Antennas, Electromagnectis.